\documentclass[11pt]{revtex4-2}

\usepackage{hyperref}
\usepackage{amsmath,amsfonts,amssymb}
\hypersetup{dvips,dvipdfm,colorlinks=true,urlcolor=magenta,filecolor=magenta,linktoc=page,citecolor=red,linkcolor=blue,bookmarks=true}
\usepackage{graphicx,epstopdf}

\begin{document}

	\title{Analytic Solutions and Observational support: A study of $f(R,T)$ gravity with $f(R,T)=R+h(T)$}
	
	\author{Akash Bose$^1$\footnote {bose.akash13@gmail.com}}
	\author{Gopal Sardar$^1$\footnote {gopalsardar.87@gmail.com}}
	\author{Subenoy Chakraborty$^1$\footnote {schakraborty.math@gmail.com}}
	\affiliation{$^1$Department of Mathematics, Jadavpur University, Kolkata-700032, West Bengal, India}
	
	\begin{abstract}
		The present work deals with cosmological solutions in $f(R,T)$ gravity theory for perfect fluid with constant equation of state ($\omega$). For a viable cosmological solution $\omega$ is restricted to $\omega<\dfrac{1}{3}$. Also depending on the sign of an arbitrary constant one has two possible solutions: a finite universe model and an ever expanding model of the universe. Field theoretic description of the model and possibility of a ghost scalar field has been studied. Also, an equivalence with modified Chaplygin gas has been shown. Finally, from the observational data it has been concluded that this model represents an evolution of the universe from matter dominated phase to present accelerating phase.
		
	\end{abstract}
	
	\maketitle
	
	\section{Introduction}

	The standard cosmology has been going through an embracing phase due to a series of observational evidence \cite{Riess:1998cb,Perlmutter:1998np,Spergel:2003cb,Tegmark:2003ud,Eisenstein:2005su} for the last two decades. The present accelerated era of expansion as predicted by the observational data has tension with the theoretical prediction of standard cosmology (i.e. LCDM model) (for details one may refer to the recent works for example Snowmass reviews \cite{Cabass:2022avo}).
	  So a group of cosmologists are trying to accommodate this observational fact by considering some exotic matter (known as dark energy) in the framework of Einstein gravity. There are various choices for the dark energy models namely phantom scalar field \cite{Caldwell:1999ew},  quintessence scalar field \cite{Carroll:1998zi,Brax:1999gp}, K-essence scalar field \cite{Armendariz-Picon:1999hyi,Chiba:1999ka,Armendariz-Picon:2000ulo}, quintom scalar field \cite{Elizalde:2004mq,Feng:2004ad,Cai:2009zp}, tachyon field \cite{Padmanabhan:2002cp,Padmanabhan:2002sh}, modified Chaplygin gas \cite{Kamenshchik:2001cp,Bilic:2001cg,Bento:2002ps,Debnath:2004cd} and so on. Alternatively another group of cosmologists are trying to have a modification of gravity theory so that the mismatch between theory and observation can be overcome. The present work is an attempt considering the second view of cosmologists.
	
	To address tensions in the present expansion rate of the Universe one natural way to address this issue is by modifying gravity theory and also there is another possibility of modifying the standard model of particle physics through dark matter and other approaches. In the context of Einstein gravity, a natural replacement of the Einstein-Hilbert action is by an arbitrary function of $R$. This modified gravity theory is the well known $f(R)$ gravity theory. This modified gravity not only explains the late time cosmic acceleration \cite{Carroll:2003wy} but also satisfies local gravitational tests \cite{Nojiri:2007as,Cognola:2007zu,Elizalde:2010ts,Nojiri:2010wj,Sotiriou:2008rp}. In the recent past this modified gravity theory has been extended by Harko \textit{et al.} \cite{Harko:2011kv} by choosing the Lagrangian density as an arbitrary function $f(R,T)$ where $T$ is the trace of the energy-momentum tensor. The justification of introducing the matter part in the gravity Lagrangian is due to Quantum effect (known as conformal anomaly). Further due to the coupling in matter and geometry the gravity model obviously depends on the source term and consequently the test particles do not follow the geodesic path (as there is a hypothetical force term perpendicular to the four velocity). Due to the highly complicated form of the field equation there is a simple choice of $f(R,T)$ in an unorthodox manner $(f(R,T)=R+h(T))$ \cite{Chakraborty:2012kj}, keeping in mind the path of the test particle will be along the geodesic. Further this particular choice of $f(R,T)$ is not possible for electromagnetic fields while if we consider the matter as a perfect fluid with constant equation of state then $h(T)$ turns out to be a power law in $T$ where the power of $T$ depends on the equation of state parameter. In the present work, our motivation is to justify the above simple but unusual choice of $f(R,T)$ from the observational point of view. Also we shall make some cosmological consequences of the present model. The plan of the paper is as follows: Section \ref{s2} gives an overview of $f(R,T)$ gravity theory, particularly in the context of present choice. Section \ref{s3} describes a detailed analysis of the cosmological solution both from theoretical and observational point of view. Section \ref{s4} shows in detail, the nature of the scalar field considering the field theoretic description of the model. Section \ref{s5} shows an equivalence of the perfect fluid in $f(R,T)$ gravity theory and the standard modified chaplygin gas model. The numerical analysis with observation constraint has been presented in section \ref{s6}. Finally a brief summary of the whole work has been given in section \ref{s7}.
	\section{$f(R,T)$ gravity theory: A brief description}\label{s2}
	In this gravity theory, 
	the complete action can be written as \cite{Harko:2011kv}
	\begin{equation}\label{eq1}
	\mathcal{A}=\int\left[\frac{1}{16\pi}f(R,T)+\mathcal{L}_m\right]\sqrt{-g}d^4x
	\end{equation}
	
	where $T=T_{\mu\nu}g^{\mu\nu}$ is the trace of the energy-momentum tensor $T_{\mu\nu}$, obtained from the matter Lagrangian density as \cite{Landau:1982dva}
\begin{equation}
T_{\mu\nu}=-\frac{2}{\sqrt{-g}}\frac{\delta(\sqrt{-g}\mathcal{L}_m)}{\delta g^{\mu\nu}}
\end{equation}
	
	Further, if $\mathcal{L}_m$ depends only on $g_{\mu\nu}$ but not its derivatives, then the above  form for $T_{\mu\nu}$ simplifies to 
	\begin{equation}
	T_{\mu\nu}=g_{\mu\nu}\mathcal{L}_m-2\frac{\partial\mathcal{L}_m}{\partial g^{\mu\nu}}.
	\end{equation}
	
Now the variation of the action (\ref{eq1}) gives the field equations as \cite{Harko:2011kv}
%
	\begin{equation}\label{eq6a}
	f_RR_{\mu\nu}+\left(g_{\mu\nu}\square-\nabla_\mu\nabla_\nu\right)f_R-\frac{1}{2}g_{\mu\nu}f(R,T)=8\pi T_{\mu\nu}-\left(T_{\mu\nu}+\Theta_{\mu\nu}\right)f_T
	\end{equation}
	with
	\begin{eqnarray}
	\Theta_{\mu\nu}=g^{\alpha\beta}\frac{\delta T_{\alpha\beta}}{\delta g^{\mu\nu}}=-2T_{\mu\nu}+g_{\mu\nu}\mathcal{L}_m-2g^{\alpha\beta}\frac{\partial^2\mathcal{L}_m}{\partial g^{\mu\nu}\partial g^{\alpha\beta}},\nonumber\\
	f_R=\dfrac{\partial f(R,T)}{\partial R},\ \ \ \ \ \ \  f_T=\dfrac{\partial f(R,T)}{\partial T}.\ \ \ \ \ \ \ \ \nonumber
	\end{eqnarray}
	
	One should note that field equations for $f(R)$ gravity can be recovered from equation (\ref{eq6a}) if $f(R,T)$ is replaced by $f(R)$. Further, one may recover GR if $f(R,T)=R$ while $\Lambda$CDM model will be recovered if $R+2\Lambda$ ($\Lambda$, a cosmological constant) with matter in the form of dust i.e. $L_m=\rho$.
	
	In the present work, a particular choice namely $f(R,T)=R+h(T)$ is considered so that the field equation (\ref{eq6a}) simplifies to \cite{Chakraborty:2012kj}
	\begin{equation}\label{eq8a}
	G_{\mu\nu}=8\pi T_{\mu\nu}-h'(T)(T_{\mu\nu}+\Theta_{\mu\nu})+\frac{1}{2}h(T)g_{\mu\nu}.
	\end{equation}
	
	If divergence of the above equation (\ref{eq8a}) is considered with conservation of energy-momentum tensor (i.e, $\nabla_\mu T^\mu_\nu=0$) then one obtains
	\begin{equation}\label{eq9a}
	(T_{\mu\nu}+\Theta_{\mu\nu})\nabla^\mu h'(T)+h'(T)\nabla^\mu\Theta_{\mu\nu}+\frac{1}{2}g_{\mu\nu}\nabla^\mu h(T)=0
	\end{equation}
	
	Thus it is easy to see that $h(T)$ is not arbitrary, rather it depends on the choice of the matter field. It is to be mentioned that for this choice of $f(R,T)$, it is not possible to consider the electromagnetic field as the matter field.
	
	In the context of perfect fluid (which we shall consider in the following sections), the energy momentum tensor has the following form
	\begin{equation}
	T_{\mu\nu}=(\rho+p)u_\mu u_\nu-pg_{\mu\nu}
	\end{equation}
	with matter Lagrangian as $\mathcal{L}_m=-p$. Here $\rho$ and $p$ are the energy density and thermodynamic pressure of the perfect fluid with restrictions on the four velocity $u^\mu$ as
	$$u_\mu u^\mu=1~;~~u^\mu\nabla_\nu u_\mu=0.$$
	
	The symmetric (0,2) tensor $\Theta_{\mu\nu}$ simplifies to
	\begin{equation}
	\Theta_{\mu\nu}=-2T_{\mu\nu}-pg_{\mu\nu}.
	\end{equation}
	
	Using this form of $\Theta_{\mu\nu}$ into equation (\ref{eq9a}) one obtains
	\begin{equation}
	(T_{\mu\nu}+pg_{\mu\nu})\nabla^\mu h'(T)+h'(T)g_{\mu\nu}\nabla^\mu p+\frac{1}{2}g_{\mu\nu}\nabla^\mu h(T)=0.
	\end{equation}
	
%
%
%
%
	Moreover, if the perfect fluid is assumed to have barotropic equation of state: $p=\omega\rho$, $\omega$ a constant, then in cosmological context for flat FLRW space-time $h(T)$ has the following simple power law form as \cite{Chakraborty:2012kj}
	\begin{equation}\label{eq13}
	h(T)=h_0T^\alpha
	\end{equation}
	
	with $\alpha=\dfrac{1+3\omega}{2(1+\omega)}$, $h_0$ is integration constant and $\omega\neq-1,\pm\dfrac{1}{3}$.

\section{Cosmology in $f(R,T)$ gravity theory}\label{s3}
	
	In the background of homogeneous and isotropic space-time geometry and with equation (\ref{eq13}) as the choice for $h(T)$, the modified field equations in $f(R,T)$ gravity theory can be written as 
	\begin{eqnarray}
	3H^2=\rho+h_0(1-3\omega)^{\alpha-1}\rho^\alpha\label{eq14}\\
	\mbox{and~~~}2\dot{H}+3H^2=-p+\dfrac{1}{2}h_0(1-3\omega)^\alpha\rho^\alpha\label{eq15}
	\end{eqnarray}	
with 	matter field conservation equation
	\begin{equation}
	\dot{\rho}+3H(p+\rho)=0.\label{eq16a}
	\end{equation}

Due to constant equation of state (i.e, $p=\omega\rho$, $\omega$ a constant) equation (\ref{eq16a}) can be integrated to give
\begin{equation}
\rho=\rho_0(1+z)^{3(1+\omega)}~,~~~~\rho_0,~\mbox{a constant of integration}\label{eq17a}
\end{equation}	
where the redshift parameter $z$ is defined as $\dfrac{a_0}{a}=(1+z)$.
	Using this solution for $\rho$ in the 1st modified Friedmann equation (\ref{eq14}) and solving for the redshift parameter one gets
	\begin{equation}
	t=t_0+\frac{2}{\sqrt{3\rho_0}}\dfrac{(1+z)^{-\dfrac{3(1+\omega)}{2}}}{(1+\omega)}{}~_2F_1\left(\dfrac{1}{2},\beta-1,\beta,x\right)
	\end{equation} 
	where $\beta=\dfrac{2}{1-\omega},~x=-h_0\left[\rho_0(1-3\omega)\right]^{\dfrac{\omega-1}{2(1+\omega)}} (1+z)^{\dfrac{3(\omega-1)}{2}}$.	
	$t_0$ is a constant of integration and ${}_2F_1$ is the usual confluent hypergeometric function. Also the Hubble parameter and the measure of acceleration can be expressed in terms of the redshift parameter as
	\begin{eqnarray}
	H^2&=&\dfrac{\rho_0}{3}(1+z)^{3(1+\omega)}(1-x)\nonumber\\
	&=&\dfrac{\rho_0}{3}\left[(1+z)^{3(1+\omega)}+h_0(1-3\omega)^{\dfrac{\omega-1}{2(1+\omega)}}\rho_0^{\dfrac{\omega-1}{2(1+\omega)}}(1+z)^{\dfrac{3(1-\omega)}{2}}\right]
	\end{eqnarray}
and	\begin{equation}
	\dfrac{\ddot{a}}{a}=-\dfrac{\rho_0}{6}(1+3\omega)(1+z)^{3(1+\omega)}+\dfrac{h_0}{12}(1-9\omega)(1-3\omega)^{\dfrac{\omega-1}{2(1+\omega)}}\rho_0^{\dfrac{1+3\omega}{2(1+\omega)}}(1+z)^{\dfrac{3(1+3\omega)}{2}}.
	\end{equation}
	
	 Note that for a viable cosmological solution, one must have $\omega<\dfrac{1}{3}$. Also if $h_0<0$, one can see that the Hubble parameter vanishes for a certain value of scale factor and it does not exist after that. Hence the Universe is bounded. So to obtain  an unbounded cosmological solution, one must consider $h_0>0$ and consequently $\ddot{a}\gtrless0$ for different choices of $h_0$, $\rho_0$ and $\omega$. For acceleration or deceleration of the model one must have the following conditions (for unbounded universe i.e, $h_0>0$) 
	 \begin{align}
	 \dfrac{1}{9}\leq\omega<\dfrac{1}{3} ~~:&~~\mbox{decelerating model~} (\ddot{a}<0)\nonumber \\ -\dfrac{1}{3}<\omega<\dfrac{1}{9}~~:&~~ \mbox{no definite conclusion, depends on~} \dfrac{\rho_0}{h_0}\nonumber\\
	 \omega\leq-\dfrac{1}{3}~~:&~~ \mbox{accelerating model~} (\ddot{a}>0)\nonumber
	 \end{align}

Further, the modified Friedmann equations (\ref{eq14}) and (\ref{eq15}) in $f(R,T)$ gravity theory can be equivalent to evolution equations in Einstein gravity with non-interacting two fluid system -- both of which are perfect fluid. The first one is the usual matter field with constant equation of state $\omega$ while the second effective perfect fluid has energy density and thermodynamic pressure as $\rho_e=h_0(1-3\omega)^{\alpha-1}\rho^\alpha$ and $p_e=-\dfrac{1}{2}h_0(1-3\omega)^\alpha\rho^\alpha$. So the effective perfect fluid has barotropic equation of state $\omega_e=-\dfrac{1}{2}(1-3\omega)$. Further, the modified field equation can be written as Einstein field equation with a perfect fluid as
\begin{equation}
3H^2=\rho_T,\mbox{~and~}2\dot{H}=-\rho_T(1+\omega_T)
\end{equation}
where the variable equation of state parameter is given by
\begin{equation}
	\omega_T=\frac{\omega-\dfrac{1}{2}h_0(1-3\omega)^{\frac{3\omega+1}{2(\omega+1)}}\rho_0^{\frac{\omega-1}{2(\omega+1)}}(1+z)^{\frac{3}{2}(\omega-1)}}{1+h_0(1-3\omega)^{\frac{\omega-1}{2 (\omega+1)}}\rho_0^{\frac{\omega-1}{2(\omega+1)}}(1+z)^{\frac{3}{2}(\omega-1)}}.
\end{equation}

\begin{figure}[h]
	\begin{minipage}{0.38\textwidth}
		\includegraphics[height=.25\textheight]{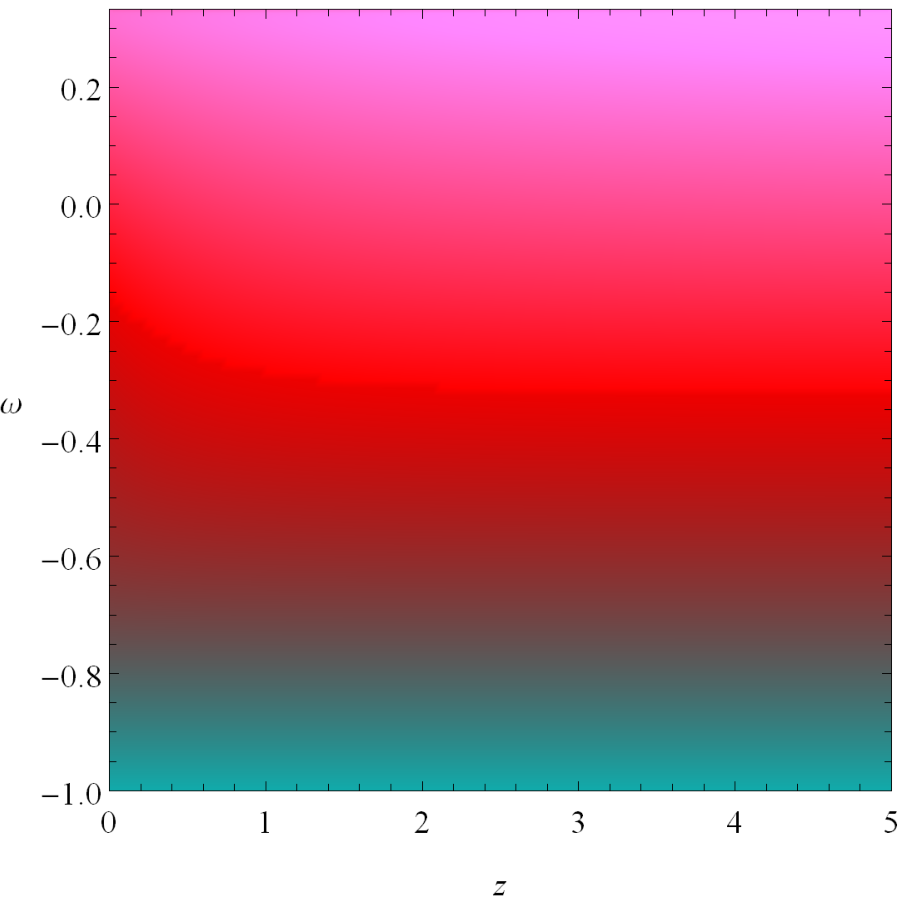}
	\end{minipage}
	\begin{minipage}{0.05\textwidth}
		\includegraphics[height=.2\textheight]{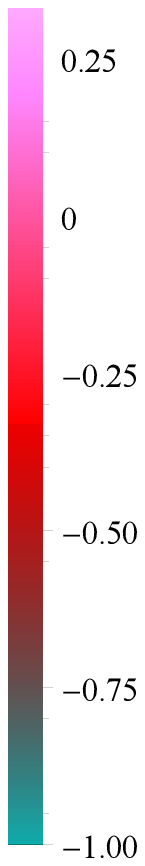}
	\end{minipage}
\caption{Variation of $\omega_T$ with respect to $\omega$ and $z$ with $\rho_0=2$ and $h_0=0.9$. }
	\label{f4}
\end{figure}

The variation of $\omega_T$ with the variation of $\omega$ and redshift parameter $z$ has been  shown in FIG. \ref{f4}.

The present $f(R,T)$ gravity model or equivalently two fluid Einstein gravity model can not describe warm inflationary scenario for the following two reasons: (i) There is no interaction between the two fluids and as a result dissipative or friction term is absent in the matter evolution equations (ii) Neither the matter fluid nor the effective fluid can have radiation equation of state. Moreover, the present cosmological model is equilibrium thermodynamical prescription due to non-existence of any dissipative pressure in both the fluids. Further, the present effective Einstein gravity with two fluids, the effective fluid will always be exotic in nature (i.e, DE) provided $\omega<\dfrac{1}{9}$ while it will be a normal fluid for $\dfrac{1}{9}<\omega<\dfrac{1}{3}$.\begin{figure}[h]
	\includegraphics[scale=0.7]{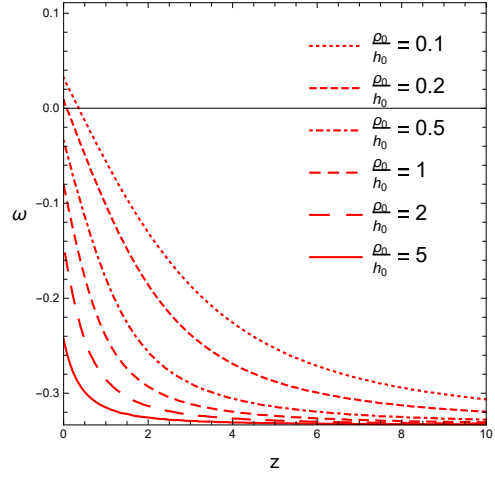}
	\caption{ $\ddot{a}=0$ is plotted in $\omega-z$ plane for different choices of $\dfrac{\rho_0}{h_0}$.}
	\label{f1}
\end{figure} Also from FIG. \ref{f1}, one may conclude that the present $f(R,T)$ model can describe the evolution of the universe from the decelerating phase to the present accelerating era for suitable choices of $\omega$ and $\dfrac{\rho_0}{h_0}$. Lastly, the cosmological parameters namely the scale factor, Hubble parameter and acceleration parameter are shown graphically in FIG. \ref{f2} for various choices of $\omega$.
\begin{figure}[h]
	\begin{minipage}{0.3\textwidth}		\includegraphics[height=.15\textheight]{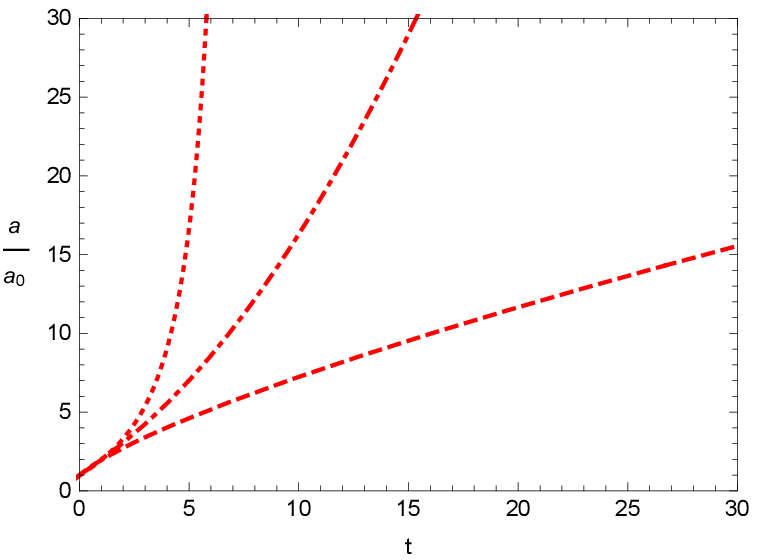}
	\end{minipage}
	\begin{minipage}{0.3\textwidth}
		\includegraphics[height=.15\textheight]{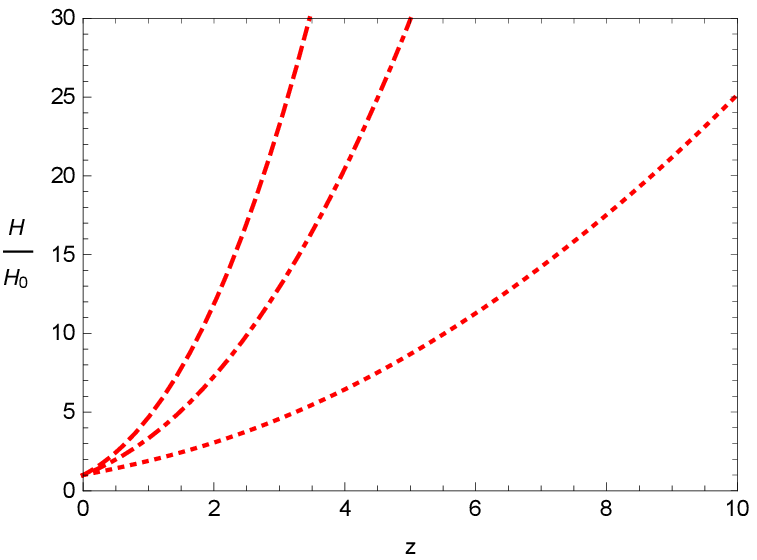}
	\end{minipage}
	\begin{minipage}{.3\textwidth}
		\includegraphics[height=.15\textheight]{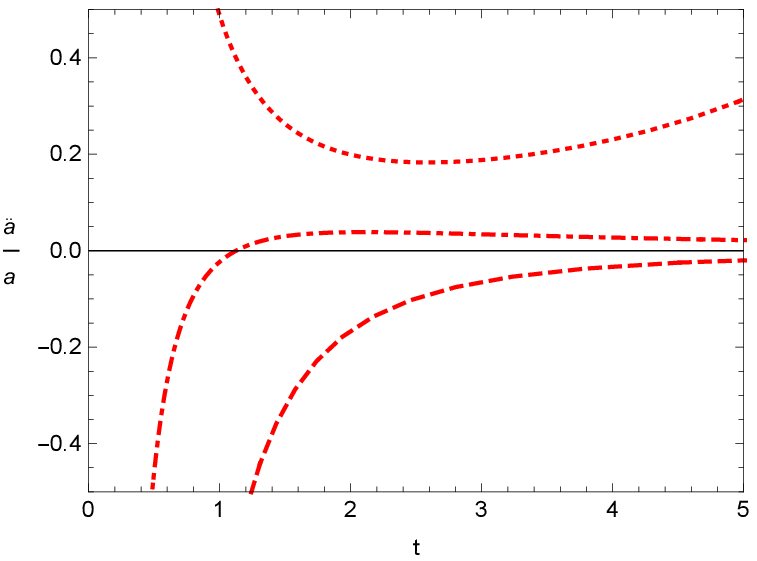}
	\end{minipage}
\caption{Scale factor, Hubble parameter, acceleration of the Universe are plotted for three different choices of $\omega$, $\omega=.25$ (dashed), 	$\omega=-.1$ (dotdashed) and $\omega=-.5$ (dotted) with $\rho_0=2$ and $h_0=0.9$. }
\label{f2}
\end{figure}

 \section{Field theoretic description}\label{s4}
 To describe the present $f(R,T)$ cosmological model from field theoretical point of view, a scalar field $\phi$ having self-interacting potential $V(\phi)$ is introduced to describe the effective fluid. So the 
  energy density $\rho_\phi$ and pressure $p_\phi$ of the scalar field are given by :
\begin{eqnarray}
\rho_\phi&=&\frac{\dot{\phi}^2}{2}+V(\phi)=h_0(1-3\omega)^{\alpha-1}\rho^\alpha\\
p_\phi&=&\frac{\dot{\phi}^2}{2}-V(\phi)=-\dfrac{1}{2}h_0(1-3\omega)^\alpha\rho^\alpha
\end{eqnarray}
i.e,
\begin{equation}
\dot{\phi}^2=\dfrac{1}{2}h_0(1-3\omega)^{\alpha-1}(1+3\omega)\rho^\alpha~~;~~~
V(\phi)=\dfrac{3}{4}h_0(1-3\omega)^{\alpha-1}(1-\omega)\rho^\alpha\label{eq17}.
\end{equation}

Using the solution for $\rho$ from equation (\ref{eq17a}) into equation (\ref{eq17}) and integrating $\phi$ has the explicit solution
\begin{equation}
\phi=\phi_0\sinh^{-1}\left(\mu a^r\right)
\end{equation}
and the potential function takes the form
\begin{equation}
V(\phi)=V_0a^y
\end{equation}
or in term of the scalar field
\begin{equation}
V(\phi)=V_1\left[\sinh\left(\delta \phi\right)\right]^s
\end{equation}
where $\phi_0=\dfrac{2\sqrt{\frac{2(1+3\omega)}{3}}}{1-\omega}$, $\mu=\sqrt{h_0}\left[(1-3\omega)\rho_0\right]^{\frac{\alpha-1}{2}}$, $r=\dfrac{3}{4}(1-\omega)$, $V_0=\dfrac{3}{4}h_0(1-3\omega)^{\alpha-1}(1-\omega)\rho_0^\alpha$, $y=-\dfrac{3}{2}(1+3\omega)$, $V_1=\dfrac{3(1-\omega)h_0^{\frac{2(1+\omega)}{1-\omega}}}{4 (1-3\omega)}$, $\delta=\dfrac{1-\omega}{2}\sqrt{\dfrac{3}{2 (1+3 \omega)}}$, $s=\dfrac{2(1+3\omega)}{\omega-1}$.   
%

\begin{figure}[h]
	\includegraphics[height=.2\textheight,width=.2\textheight]{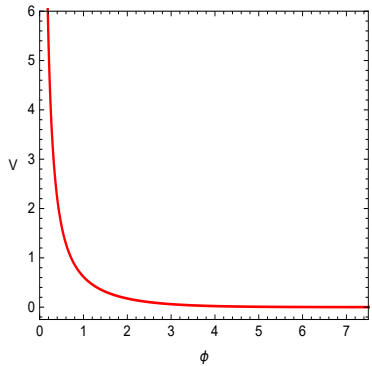}
	\caption{$V-\phi$ plot: $h_0=0.9,$ $\rho_0=2$, $\omega=-0.1$}\label{f3}
\end{figure}

The variation of the potential over scalar field has been presented in FIG. \ref{f3}.

Further, eliminating $\rho$ between the equations in (\ref{eq17}) one obtains
\begin{eqnarray}
	\frac{\dot{\phi}^2}{V(\phi)}&=&\frac{2}{3}\frac{(1+3\omega)}{(1-\omega)}\nonumber\\
	\mbox{i.e.,~}\int\frac{d\phi}{\sqrt{V(\phi)}}&=&\sqrt{\frac{2(1+3\omega)}{3(1-\omega)}}(t-t_0)
\end{eqnarray}

So the above field theoretic description is possible for $-\dfrac{1}{3}<\omega<1$. In particular, for some known potential we have the explicit form of the scalar field as
\begin{eqnarray}
	\mathrm{(i)}~ V(\phi)=V_0\phi^{-2(n-1)}&,& \phi^n=n\sqrt{\frac{2V_0(1+3\omega)}{3(1-\omega)}}(t-t_0)\nonumber\\
	\mathrm{(ii)}~ V(\phi)=V_0 ~\mathrm{sech}^2\phi~&,& \phi=\sinh^{-1}\left[\sqrt{\frac{2V_0(1+3\omega)}{3(1-\omega)}}(t-t_0)\right]\nonumber
\end{eqnarray}

From the expression of $\dot{\phi}^2$ in equation (\ref{eq17}), it is easy to see that the equivalent scalar field in field theoretic description will be a real scalar field provided either (i) $h_0>0$ and $\omega>-\dfrac{1}{3}$ i.e, normal fluid in an unbounded universe, or (ii) $h_0<0$ and $\omega<-\dfrac{1}{3}$ i.e, exotic fluid in bounded universe.

Thus an unbounded model of the universe will be ever accelerating if the equivalent scalar field is of ghost nature while the unbounded universe will be ever decelerating or undergoes a transition from deceleration to acceleration for equivalent scalar field to be real.  
\section{An equivalent notion of modified Chaplygin gas and perfect fluid in $f(R,T)$ gravity theory}\label{s5}

This section presents a nice equivalence between the $f(R,T)$ gravity theory with the typical choice of the $f(R,T)$ function (in the present work) and Einstein gravity. The modified Einstein field equations in $f(R,T)$ gravity theory for perfect fluid in FLRW space-time geometry are given by equations (\ref{eq14}) and (\ref{eq15}) with $\omega=\dfrac{p}{\rho}$ as the equation of state parameter. Now suppose there exist a modified Chaplygin gas (MCG) having equation of state 
\begin{equation}\label{eq27}
p=\gamma\rho-\Gamma\rho^\alpha
\end{equation}
with $\gamma$, $\Gamma$ and $\alpha$ are constant parameters of the Chaplygin gas given by$$\gamma=\dfrac{1}{2}(1-3\omega),~\Gamma=\dfrac{1}{2}|h_0|(1-3\omega)^\alpha.$$

 If the above modified field equations are considered as equivalent Einstein field equations then one may  write
%
\begin{eqnarray}
3H^2&=~\rho_e
&=\frac{p}{\gamma}\nonumber\\
2\dot{H}+3H^2&=-p_e
&=-\gamma\rho
\end{eqnarray}


Thus the effective single fluid in Einstein gravity has the equation of state $p_e=\omega_e\rho_e$ where $\omega_e=\dfrac{\gamma^2}{\omega}$, a constant. So we may conclude that a MCG fluid in $f(R,T)$ gravity theory is equivalent to a perfect fluid in Einstein gravity with constant equation of state $\dfrac{\gamma^2}{\omega}$. Therefore, the present choice of the $f(R,T)$ function shows that the equivalence between the modified gravity theory and Einstein gravity is just an exchange of a perfect fluid with constant equation of state to the well known modified chaplygin gas.

\section{Numerical Analysis and Observational Constraint}\label{s6}

Our aim here is to constraint on the cosmological parameters analyzing the observational data sets. In order to do so, the pressure of the matter component and effective fluid can be rewritten as
\begin{eqnarray}
	p_m\equiv p=\omega\rho&\equiv&w0_{b}\rho_m\label{eq31}\\
	p_{fld}\equiv p_e=-\frac{1}{2}(1-3\omega)\rho_e&\equiv&-\frac{1}{2}(1-3w0_{b})\rho_{fld}
\end{eqnarray}
where the corresponding energy density for two  components can be rewritten as
\begin{eqnarray}
	\rho_m\equiv \rho=\rho_0 (1+z)^{3(1+\omega)}&\equiv&\Omega_{m}H0^2 (1+z)^{3(1+w0_{b})}\\
	\rho_{fld}\equiv \rho_e=h_0(1-3\omega)^{\alpha-1}\rho^\alpha&\equiv&\Omega0_{fld}H0^2 (1+z)^{\frac{3}{2}(1+3w0_{b})}\label{eq34}
\end{eqnarray}
 and  the public version of the CLASS Boltzmann code has been modified to include the dark energy sector as effective fluid  and  for baryonic matter and corresponding cold dark matter (Eqns. (\ref{eq31})-(\ref{eq34})). The MCMC code Montepython3.5 \cite{Brinckmann:2018cvx} has been used to estimate the relevant cosmological parameters.

In order to analysis and make a comparison, we use the  
cosmological datasets as dataset I (Pantheon \cite{Pan-STARRS1:2017jku}, BAO (BOSS DR12 \cite{BOSS:2016wmc}, $SMALLZ-2014$ \cite{Ross:2014qpa}) and HST \cite{Riess:2011}) and dataset II (Pantheon \cite{Pan-STARRS1:2017jku}, HST \cite{Riess:2011}). In both cases, a PLANCK18 prior  has been
imposed. 

We have made the choice of flat priors on the base cosmological parameters as follows: the baryon density $100\omega_b = [1.9, 2.5]$; cold dark matter density $\omega_{cdm} = [0.095, 0.145]$; Hubble parameter $H0 =
[60, 80] km s^{-1}Mpc^{-1} $ and a wide range of flat prior has been chosen for $w0_b=[-1, 1]$.

\begin{figure}
\centering
\includegraphics[scale=0.35]{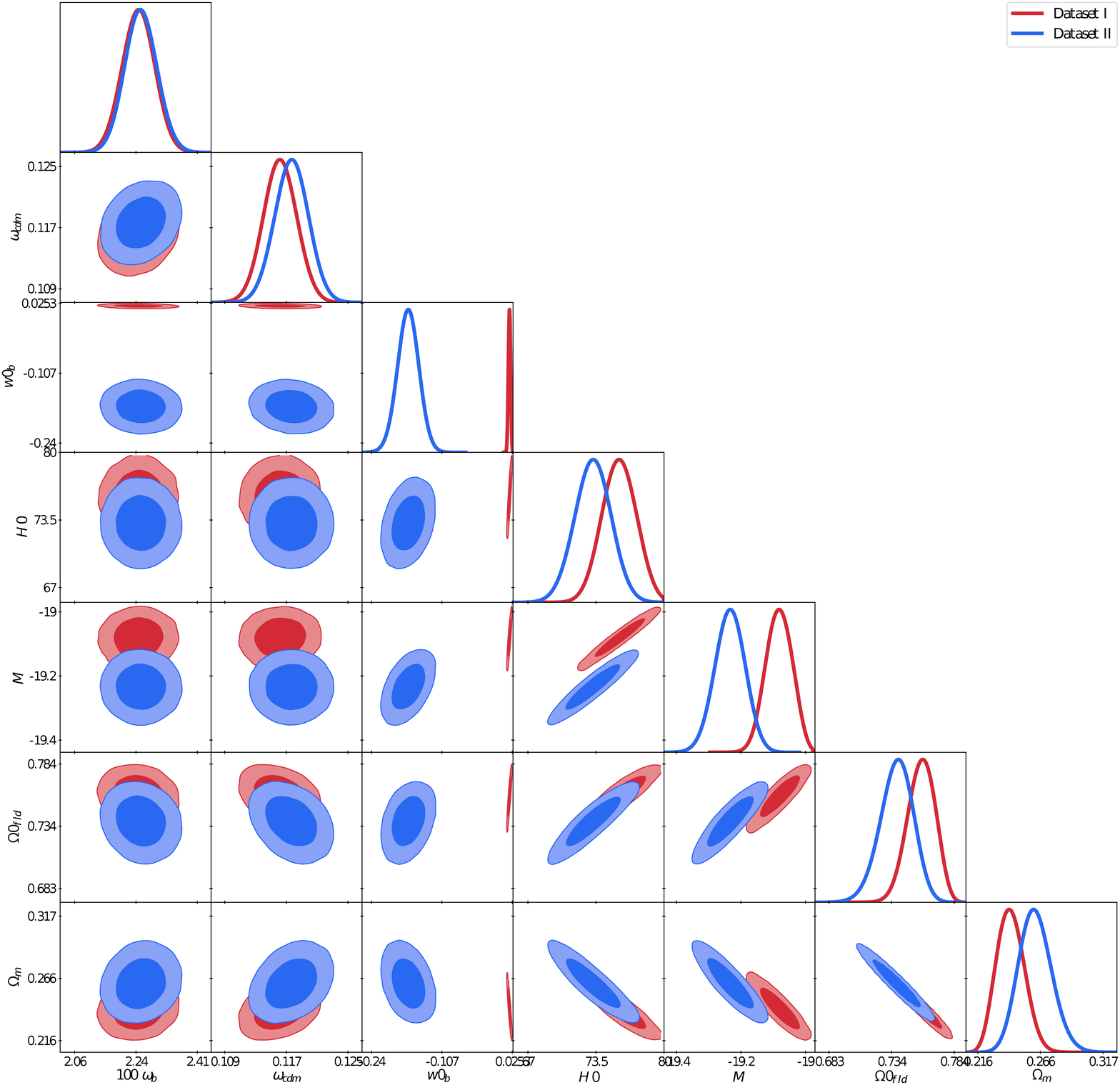}
\caption{Posterior distribution for the cosmological parameters using the 
datasets I and datasets II. In the both cases a PLANCK18 prior has been 
considered.}\label{f5} 
\end{figure}

In the Table. \ref{table1} we have enlisted the  constraints on the various cosmological parameters and in the Fig. \ref{f5}, we have shown the posterior distribution of those parameters.
 
\begin{table}[h]
\begin{tabular}{|c|c|c|c|c|}
\hline
 & \multicolumn{2}{c|}{Dataset I} & \multicolumn{2}{c|}{Dataset II} \\
\cline{1-5}
Param & best-fit & mean$\pm\sigma$ & best-fit & mean$\pm\sigma$ \\ \hline 
$100~\omega{}_{b }$ &$2.244$ & $2.242_{-0.046}^{+0.045}$ & $2.252$ & $2.249_{-0.046}^{+0.045}$ \\ \hline
$\omega{}_{cdm }$ &$0.1158$ & $0.1162_{-0.0023}^{+0.0023}$ & $0.118$ & $0.1177_{-0.0023}^{+0.0023}$ \\ \hline
$w0_{b }$ &$0.02035$ & $0.02012_{-0.002}^{+0.0022}$ &$-0.1722$ & $-0.1706_{-0.021}^{+0.021}$\\ \hline
$H0$ &$75.77$ & $75.71_{-1.6}^{+1.7}$ &$73.11$ & $73.18_{-1.8}^{+1.7}$ \\ \hline
$M$ &$-19.06$ & $-19.07_{-0.044}^{+0.049}$ &$-19.24$ & $-19.24_{-0.054}^{+0.053}$  \\ \hline
$\Omega{} 0_{fld }$ &$0.7591$ & $0.7577_{-0.01}^{+0.012}$ & $0.737$ & $0.7376_{-0.013}^{+0.014}$\\ \hline
$\Omega{}_{m }$ &$0.2407$ & $0.2422_{-0.012}^{+0.01}$ &$0.2629$ & $0.2623_{-0.014}^{+0.013}$\\ 
\hline 
$\chi^2_{\min}$&\multicolumn{2}{c|}{1130} & \multicolumn{2}{c|}{1029}\\
\hline
 \end{tabular}
\caption{}\label{table1} 
\end{table}
  
From the above numerical analysis and the observed data we found that the equation of state parameter turns out to be $w0_b=0.02035$ (for DataSet I), $w0_b=-0.1722$ (for DataSet II) which is consistent with the theoretical prediction $\left(-\dfrac{1}{3}<\omega<\dfrac{1}{9}\right)$ in the present work for the $f(R,T)$ modified gravity model. Moreover, the above observational data analysis shows a transition of model from decelerated era of expansion to the present accelerated expansion era as inferred by the theoretic prediction. Now using the best fit values of the parameters from TABLE \ref{table1}, $\rho_0$ and $h_0$ can be estimated as
\begin{table}[h!]
	\begin{tabular}{|c|c|c|}\hline
	Best fit&$\rho_0$&$h_0$\\\hline
	Dataset I&1381.88&98.47\\\hline
	Dataset II&1405.22&637.392\\\hline
	\end{tabular}
\end{table}

and hence the choice of $f(R,T)$ becomes $f(R,T)=R+98.47T^{0.52}$ (Dataset I) and $f(R,T)=R+637.39T^{0.29}$ (Dataset II). So one may note that after adding the BAO data (Dataset I) the parameter $w0_b$ is estimated as higher value compared to Dataset II, consequently the value of $H0$ increases for Dataset I which is consistent with FIG. 3. Further, it can be concluded that  the trace of the energy momentum tensor is more significant in the $f(R,T)$ choice when the BAO data is absent.
\section{Summary}\label{s7}

An extensive study of FLRW cosmology in $f(R,T)$ gravity has been presented in the present work with a suitable choice of the function $f(R,T)$. The matter field is chosen as perfect with constant equation of state. The modified field  equations are equivalent to non-interacting two fluid system in Einstein gravity. The effective fluid is also a perfect fluid with constant equation of state (depending on the state parameter of the given physical fluid). A possible cosmological solution for the present model has been obtained and the corresponding values of Hubble parameter and acceleration parameter are determined and their graphical representation has been shown in FIG \ref{f2}. Also the variation of equation of state parameter for the combined single fluid has been shown graphically in a 3D plot against redshift parameter and equation of state parameter for the physical fluid. Depending on the sign of a parameter ($h_0$) the present model may describe  a bounded universe or an unbounded model of the universe. Due to the non-interacting nature of the two fluid system, the model can not describe the warm inflationary scenario and the present model is in thermodynamical equilibrium configuration. Depending on the choices of equation of state parameter of the usual fluid and the ratio $\left(\dfrac{\rho_0}{h_0}\right)$ the model may describe the evolution from decelerating phase to the present accelerating era. This result is in accord with the work of M. J. S. Houndjo \cite{Houndjo:2011tu}. There is a field theoretic description of the  model with effective fluid described by a scalar field. From this field theoretic description it is found that an unbounded model of the Universe will be ever accelerating if the nature of the equivalent scalar field is of ghost type while for a real scalar field the universe will experience either ever decelerating phase or a transition from the deceleration to acceleration. Also, it is interesting to see that a MCG in $f(R,T)$ gravity is equivalent to a perfect fluid in Einstein gravity. Thus with the transition from one gravity theory to the other, the physical fluid also makes a transition to another form of fluid. Moreover, From observational point of view the parameters involved in the present model are estimated from Pantheon Data \& BAO Data. It is found that the contribution of the matter part (by the trace term T) in the model function $f(R,T)$ is dominant when BAO data is not taken into account. Finally, both from the theoretical prediction and from the observational data analysis, one may infer that the present $f(R,T)$ gravity model describes the cosmic scenario from matter dominated era to the present late time accelerating phase, it does not predict the early era of the universe.
\section*{Acknowledgement}
The authors would like to thank Nandan Roy and Supriya Pan for helping to run Montepython. The author A.B. acknowledges UGC-JRF (ID:1207/CSIRNETJUNE2019). G.S. acknowledges  UGC for Dr. D.S. Kothari Postdoctoral Fellowship (No.F.4-
2/2006 (BSR)/PH/19-20/0104) and S.C. thanks Science and Engineering Research Board (SERB), India for awarding MATRICS Research Grant support (File No.MTR/2017/000407).
  	
\end{document}